\documentstyle[prb,aps]{revtex}

\input{psfig}

\begin{document}
\title{
Comparison of local structure measurements from $c$-axis polarized XAFS 
between a film and a single crystal of YBa$_2$Cu$_3$O$_{7-\delta}$ as a function of temperature
}
\author{C.  H.  Booth, F.  Bridges}
\address{
Physics Department,
University of California,
Santa Cruz, CA 95064
}
\author{J. B. Boyce}
\address{
Xerox Palo Alto Research Center, Palo Alto, CA 94304
}
\author{T. Claeson}
\address{
Physics Department,
Chalmers Univ.  of Tech.,
S-41296 Gothenburg, Sweden
}
\date{Draft: \today}

\maketitle
\begin{abstract}
We have performed fluorescence x-ray absorption fine-structure (XAFS) 
measurements from 20-200 K on a 5000 \AA,
$c$-axis film of YBa$_2$Cu$_3$O$_{7-\delta}$ (YBCO) on MgO ($T_c$=89 K) using photons polarized perpendicular to 
the film.  The quality of the data is high out to 16 \AA$^{-1}$.
The data from 3-15.5 \AA$^{-1}$ were transformed into $r$-space
and fit to a sum of theoretical standards
out to 4.0 \AA.  These data are compared to YBCO data from
a single crystal and from a film on LaAlO$_3$ with the same $T_c$.  The main difference between
the single crystal and the film data is that 
while the single crystal data is well described by a two-site axial oxygen [O(4)] distribution,
we see no evidence for such a distribution in either thin film sample.  We place an upper limit on
the size of an axial oxygen splitting for the film on MgO at 
$\Delta_r\lesssim$0.09 \AA.  Therefore, the magnitude of the splitting is not
directly related to $T_c$.  Fits to the 
temperature dependent data from the YBCO film on MgO indicate that all 
bonds show a smooth change of their broadening factor $\sigma$, 
except the Cu-O(4) bonds, which show an increase in
$\sigma$ in the vicinity of $T_c$, followed by a decrease of the same magnitude.
Such a feature does not occur in diffraction
measurements.  Since XAFS measurements of $\sigma$
include any correlation between the atoms in a given bond, we
conclude that the O(4) position becomes less correlated with the Cu positions
near $T_c$.  Correlation measurements of these and several further near-neighbors are also reported.

\end{abstract}
\twocolumn

%
%
%

\section{Introduction}

\begin{figure}
\psfig{file=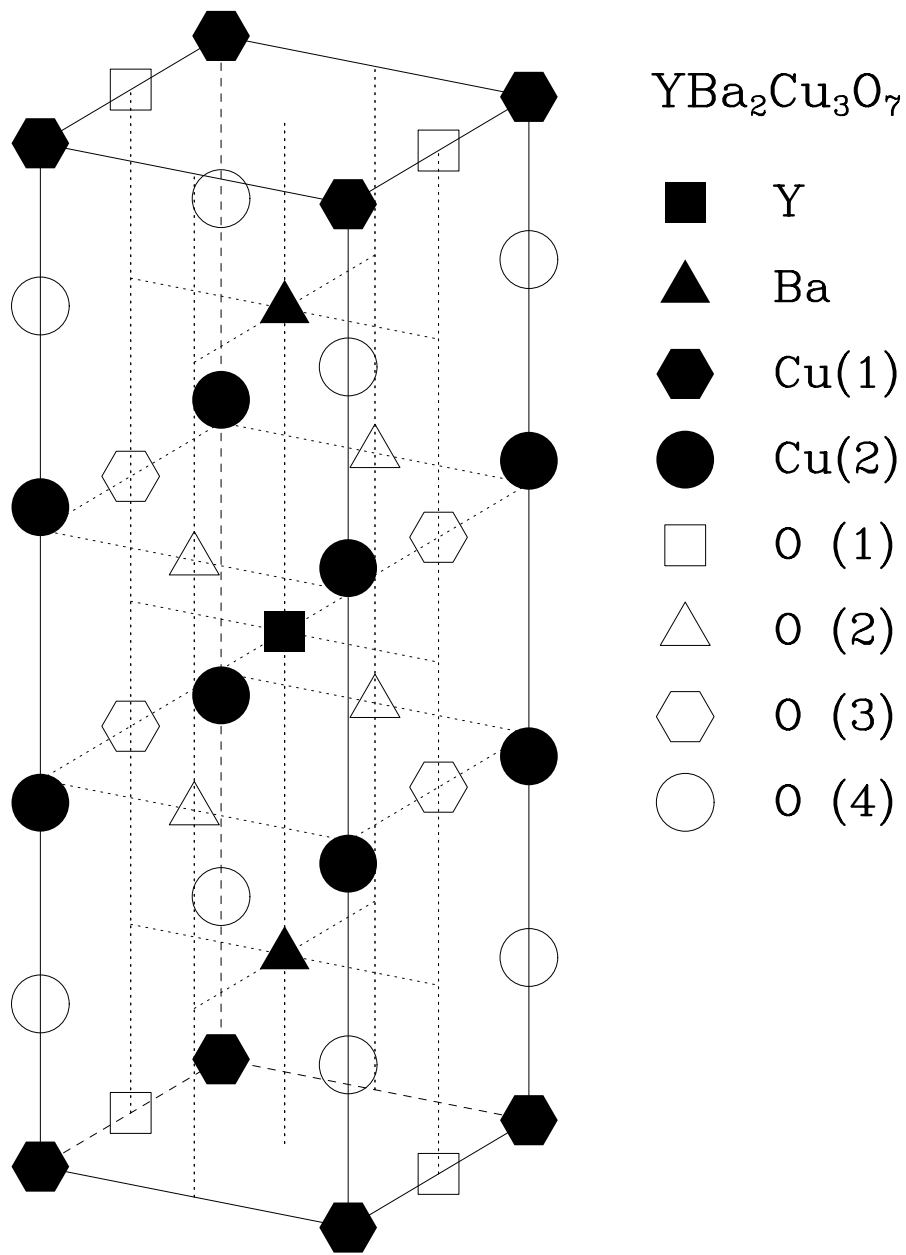,height=8in}
\caption{The crystal structure for YBa$-2$Cu$_3$O$_{7-\delta}$.  The O(2) and O(3)
oxygens are often referred to as the ``in-plane'' oxygens.  Much of
this paper is concerned with the O(4), otherwise known as the ``axial''
oxygen.
}
\label{YBCO_struc}
\end{figure}

Since the discovery of the high $T_c$ superconductors\cite{Bednorz86} there have been several
measurements of possible structural changes near the superconducting 
transition.
Some measurements were far enough above the noise that the result
was immediately convincing to the scientific community, such as Sharma {\it et
al}'s ion channeling measurements of a sudden correlation of the alignment
in the $ab$-plane
of the Cu(1), O(4) and Cu(2) atoms in ErBa$_2$Cu$_3$O$_7$\cite{Sharma89} and 
in YBa$_2$Cu$_3$O$_{7-\delta}$ (YBCO).\cite{Sharma89b} (Fig. \ref{YBCO_struc} shows 
structure and site definitions for YBCO.)
Some have relied on relatively small changes in a relatively large signal,
such as Horn {\it et al}'s\cite{Horn87} diffraction measurement of an anomaly in the 
orthorhombicity [$(b - a)/(b+a)$] of YBCO, or Mustre de Leon {\it et al}'s
x-ray-absorption fine structure (XAFS) measurement of a split O(4) position distribution ($\delta r \simeq $0.13
\AA) in YBCO that decreases somewhat or disappears ($\delta r \lesssim $0.11 \AA) near 
$T_c$.\cite{Leon90,Leon92} 
Both of these measurements were initially treated with a fair
amount of skepticism, because obtaining a reliable signal for measuring such small changes
is difficult. 
Horn {\it et al}'s result has since been confirmed both
by diffraction\cite{You91} and by capacitive dilatometry measurements\cite{Meingast91}
and consequently must be
considered real, although its relation to $T_c$ is still unclear.  On
the other hand, Mustre de Leon {\it et al}'s measurement has only been partially
confirmed by XAFS\cite{Stern93} (see below), and no other experiment has been able to 
verify the result in YBCO.

The XAFS measurement of a split O(4) site distribution has been taken as
a possible explanation for anomalous features in vibrational spectra involving
the O(4).\cite{Mihailovic93}
This distribution may also fits into current polaron models for 
electron-phonon-coupling (for instance, see Refs. \onlinecite{Emin86,Emin92,Ranninger92,Hardy92}),
In spite of these important implications, the possibly dynamic
nature of the O(4) distribution has never been verified.  In fact, it has
been suggested that the split distribution is actually due to ordered 
O(1) vacancies,\cite{Rohler94} as in the Ortho-II phase of YBCO.\cite{Zeiske91,Zeiske92b}
Part of the ambiguity in the interpretation lies in the
difficulty of the measurement itself.
The most serious problems with the XAFS measurements are pointed out by Stern
{\it et al}.\cite{Stern93}  In their attempt to confirm the Mustre de Leon {\it et al}
measurements, they tried several samples with various oxygen concentrations.
The only sample that was clearly consistent with a split distribution of
the O(4) position had the lowest $T_c$ of the samples measured (89 K).  Fits to a split distribution for
the other samples were less convincing; both the single-site fit and the
split-site fit were unsatisfactory.  The sample with the highest $T_c$ (92 K)
did not show any temperature dependence near $T_c$, at least as far as the
fits were concerned.  Since the measurement relies on
such a small change in the overall signal and the results were severely
sample dependent, the authors concluded that an independent method is necessary
to confirm the split-site character of the O(4) distribution. 
They also concluded that
the relation between the split distribution and $T_c$ must be weak and
unimportant to superconductivity since the highest $T_c$ sample showed the
least change with temperature.  This conclusion is weak, however, because
the fits to this $T_c$=92 K sample were of poor quality.

Unfortunately, no independent experiments have been able to confirm the
result in YBCO, 
although x-ray diffraction analysis has seen a much larger effect in
single crystals of 
YBa$_2$Cu$_{2.78}$O$_7$,\cite{Bordet93} pulsed neutron diffraction (using
a PDF analysis\cite{Toby90} has seen correlated displacements of the Cu
and axial oxygen in Tl-2212, and XAFS has seen axial oxygen changes near $T_c$ in
Tl-1234.\cite{Allen91}

As Stern {\it et al} have pointed out, until an independent experiment
confirms or refutes the possibly split nature of the O(4) position,
the split nature of the O(4) site will be questionable.  However, with different samples, analysis
techniques, etc. XAFS can still help clarify the issue.  
For instance, well oriented samples
are important for these XAFS measurements because the polarization of the
incoming synchrotron light can be utilized to remove the contribution to the
XAFS of the in-plane oxygens.  
Magnetically-oriented
powders were used in both the
Mustre de Leon {\it et al} and the Stern {\it et al} studies.  
Each of these absorption measurements were made in the
transmission mode.  
In the study presented here, we use a 5000 \AA{ } film
on MgO, a 4000 \AA{ } film on LaAlO$_3$, and a 17 $\mu$m (2\% Ni) single crystal 
of YBCO, all with a $T_c$ of 89 K. Each of these samples are highly 
oriented with the $c$-axis perpendicular to the sample surface.  
All measurements were made in the fluorescence mode. 
Our analysis focuses on the film on MgO and uses the other two samples for comparison.

In contrast to the results of
Stern {\it et al} and Mustre de Leon {\it et al}, the film data can be well described
by a single, harmonic O(4) distribution that shows only very subtle changes
near $T_c$.  This data is very reproducible over a wide range of temperatures,
and thus provides a null result to compare with any other results.  
On the other hand, the single crystal data shows a split O(4) 
distribution below $T_c$ consistent with previous measurements on oriented powders.
By comparing a case where the splitting is not present and one where it is,
we show that previous XAFS measurements of a split O(4) distribution are not 
relying on an unusually small signal.  

The measurements presented here also provide details of the harmonic broadening 
of the pair distribution of the atom pairs in the film on MgO.  These measurements show
interesting temperature dependences which are interpreted as changes in the
correlations between the atomic positions in a given atom pair.  Fluctuations in
the broadening parameters of the Cu(2)-O(4) pair, and to a lesser degree in the 
Cu(1)-O(4) pair in the vicinity of $T_c$ are also reported.

We begin our report by describing the experimental details in Sec.
\ref{Experimental details}.  The data analysis procedures are given in
Sec. \ref{Data Analysis}.  The isolation of the O(4) contribution to the XAFS
is presented in Sec. \ref{O4_isolate}.  Results of detailed fits to the film data are given 
in Sec. \ref{Fits}.  Consequences of observed features in the data are
discussed in Sec. \ref{Discussion} and the conclusions of this work are
summarized in Sec. \ref{Conclusions}.

\section{Experimental details}
\label{Experimental details}

The thin film samples of YBCO were prepared by B. Lairson, who used
Molecular Beam Epitaxy (MBE) to deposit the superconductor onto
a MgO and a LaAlO$_3$ substrate.  The film on MgO is estimated to be 5000 \AA{ }
thick, and the film on LaAlO$_3$ is approximately 4000 \AA{ } thick.  
X-ray diffraction measurements indicate less than 1\%
impurity phases.  The diffraction measurements also indicate that the films
are oriented with their $c$-axes perpendicular to the plane of the film with
less than 1\% $a$- or $b$-axis impurities for the film on MgO, and $<$3\%
for the film on LaAlO$_3$.  TEM pictures confirmed that films
created in the same way are $c$-axis oriented, with trace impurities
and that the crystals are twinned.
The samples have a $c$-axis lattice parameter of 11.64 \AA{ } at room temperature.
The transition temperatures were measured with the standard four-probe
technique and found to be 89 K.

We also obtained a fairly large single crystal from W. Hardy which contains
about 2\% Ni.  
This crystal is also $c$-axis oriented.  
The thickness is estimated to be $\sim$ 17 $\mu$m
by looking at the
Cu $K$-edge step in the x-ray transmission spectrum. 

X-ray fluorescence measurements were made on beamline 7-3 at the Stanford
Synchrotron Radiation Laboratory (SSRL).  The samples were placed in an
Oxford helium cryostat and oriented such that the incident x-ray beam was
striking the surface of the samples between 10-12 degrees.  Temperature was
regulated via two sensors, one in the helium chamber and one about two
inches above the sample.  Because of the geometry and later calibration
measurements, we estimate that the absolute temperature of the sample is
probably 3-5 K higher than the sample sensor was indicating (the ``nominal''
temperature).  Relative
changes in the temperature of these data should be much better and is
probably around 1 K.  All reported temperature measurements of the XAFS data 
are the nominal temperature.

The fluorescence x-rays were detected using a 13-element Ge detector.  This
detector allows for energy resolution that isolates the Cu $K_\alpha$ peak and
eliminates any
need for background fluorescences to be removed from these data.  
Count rates
in each channel were kept well below saturation levels ($<$ 20\% of saturation),
but were in any case corrected for the dead-time-constant ($\sim$ 5-10 $\mu$s) 
introduced by the energy-resolving 
peak-shaping filters.  Besides the ability
to resolve the Cu $K_\alpha$ fluorescence, the main utility in using the
Ge detector is to aid in the removal of Bragg peaks in the absorption 
spectrum.
Because the Bragg spikes occur at the incident beam energy, the energy 
resolving
power of the Ge detectors can partially discriminate against them.  However,
a large change in the fluorescence flux can bring the detectors much closer to
saturation, and thus the dead-time correction becomes more important.  If a
Bragg spike is particularly strong, the dead-time correction cannot
account for the additional flux accurately enough, and the affected data must
be removed.  Such removal is accomplished by replacing the affected data
by the normalized mean of the other channels.  In this way we can usually
generate a full absorption spectrum over the energies of interest with almost no
experimentally induced spikes or dips.  This method provides
a significant advantage over more traditional single channel ionization 
counters when one realizes
that for thin films on well-oriented substrates, or for
single crystals, Bragg spikes can affect data 
over tens of eV.

Unfortunately, features which we call
``glitches" can still arise from nonuniformities in the sample coupling to
nonuniformities in the incident beam.\cite{Bridges92a,Li94b}  Since these
occur in all the channels of the Ge detector, we must remove such features
by fitting a low (2nd or 3rd) order polynomial through neighboring 
(unaffected)
data points and replacing the data with the fit.  Fortunately, glitches often
only affect between one and four successive data points, and so this 
procedure does
not significantly alter the XAFS.

\begin{figure}
\psfig{file=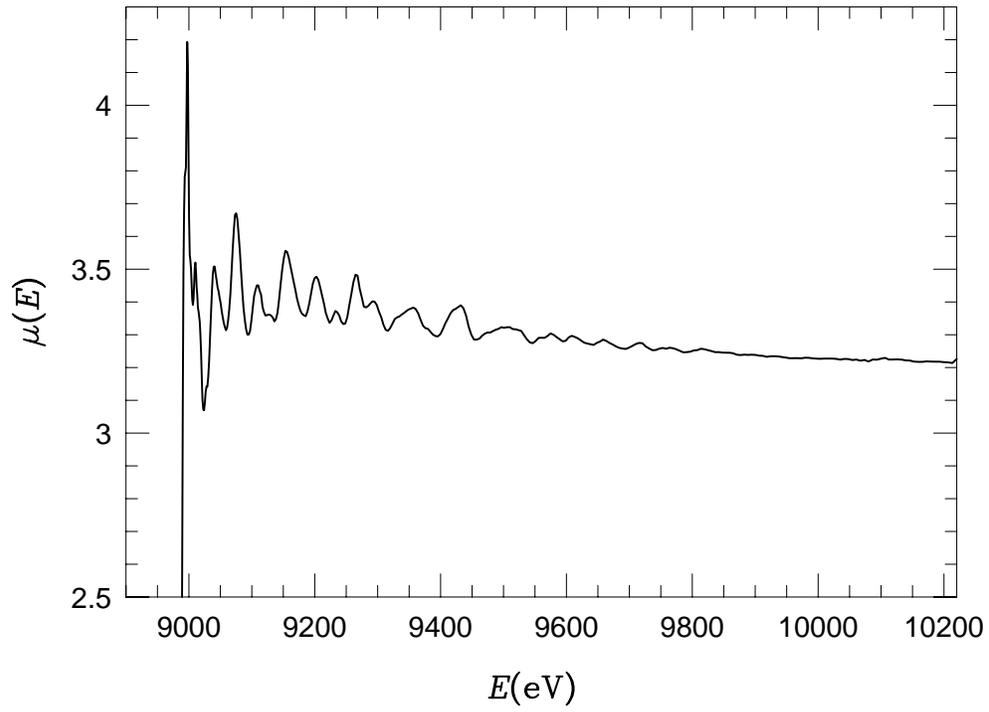,height=8in}
\caption{Full $e$-space data for a YBCO film on MgO at $T$=20 K with
the x-ray polarization parallel to the $c$-axis of the film.  These data
have been dead-time corrected and had some minor glitches removed, as described
in the text.  Data from the film on LaAlO$_3$ and the single crystal are of
comparable quality.}
\label{e-space}
\end{figure}

An example of both the film data which has been
processed as described above is displayed in Fig. \ref{e-space}.  The single
crystal data is of similar quality.

\section{Data Analysis}
\label{Data Analysis}

Our data analysis procedures have been detailed elsewhere for standard
transmission experiments.  In this work, we will 
describe
briefly the general procedure and only give details where significant
differences in the procedure arise over 
Refs. \onlinecite{Hayes82,Li95b,Bridges95b}.

\subsection{Basic XAFS}

The main features in an x-ray absorption spectrum are due to single electron
excitations, namely, the photoelectric effect.  These excitations appear
as sudden jumps in the absorption coefficient $\mu(E)$.  In a solid, the
data just above these jumps exhibit oscillatory structure which has been
named X-ray Absorption Fine Structure (XAFS).  The XAFS are isolated by
defining a function
\begin{equation}
\chi(E)\equiv\frac{\mu(E)-\mu_0(E)}{\mu_0(E)}
\label{XAFS_def}
\end{equation}
which is used just beyond the edge, normally $>$ 25 eV. $\mu_0(E)$ is a normalization
function which is the absorption due to all processes that do not include
the photoelectron backscattering effect described presently.

The oscillations in $\mu(E)$ are primarily due to an interference effect
between the outgoing photoelectron's wave function and the part of the 
photoelectron's wave function which has scattered off nearby neighbors
and returned to the absorbing atom.  As the photoelectron's wave vector is
increased, the interference is modulated with a frequency given by twice the
distance to the backscattering atom.  There are several other factors which 
must be taken into account.  The version of the ``XAFS equation'' that we use
in our data analysis is
\begin{equation}
\chi(k)=Im \sum_{i} A_i
\int_{0}^{\infty} g_i(r) 
e^{i(2 r + 2\delta_c(k) +\delta_i(k))} dr
\label{XAFS_eq}
\end{equation}
where the amplitude factor $A_i$ is given by
\begin{equation}
A_i=N_i S_{0}^2 F_i(k),
\label{A_i}
\end{equation}
$N_i$ is the number of equivalent atoms in shell $i$, $S_{0}^2$ is
the ``amplitude reduction factor'' that accounts primarily for many-body
effects that reduce the XAFS oscillation amplitude such as shake-up or 
shake-off, $F_i(k)$ is the backscattering
amplitude of the photoelectron off neighbors $i$ including a reduction
sue to the 
mean-free path of the photoelectron, $g_i(r)$ is the pair distribution
function for the absorbing and backscattering atoms, and the phase shifts
of the photoelectron from the central and the backscattering atom are
give by $\delta_c(k)$ and $\delta_i(k)$, respectively.  The distribution
function $g_i(r)$ is usually taken to be harmonic
\begin{equation}
g_i(r)=e^{\frac{-(r-R_i)^2}{2 \sigma_i^2}},
\label{harmonic}
\end{equation}
where $R_i$ is the average distance between the central and the backscattering
atoms $i$, and the broadening factor $\sigma_i$ (otherwise known as the 
Debye-Waller factor) combines the effects of 
thermal and static disorder.  If a shell $i$ contains atoms that are inequivalent, 
then $\sigma_i$ may also be used as a measure of any distortion in the shell,
even though such distortions are rarely harmonic.

In studies of well-ordered materials (such as YBCO) we expect there to be
very little difference in the bond lengths measured by XAFS compared to
diffraction.  In fact, with the exception of the O(4) site, there are many 
XAFS studies which demonstrate the strong agreement between local
structure in YBCO (and most of its relatives)
and the average structure.\cite{Boyce87,Crozier87,Boyce89}  On the other
hand, comparative measurements of the correlated Debye-Waller factors given 
by XAFS and the uncorrelated ones given by diffraction have been virtually 
ignored by most experimenters.  

Part of the reason for not making direct comparisons between XAFS and 
diffraction broadening factors is that getting the absolute broadening factor 
is not always possible.  Experimental measurements of amplitude functions have their own (usually 
unknown) broadening, and until
recently, theoretical calculations were not of high enough quality to give
reliable results.  FEFF6\cite{FEFF5,FEFF6}, written by Rehr, Zabinsky and co-workers,
has been shown to calculate accurate backscattering amplitudes and phase
shifts, and even gives reasonable values for Debye-Waller parameters for simple
materials with known Debye temperatures.\cite{Li95b,Zabinsky_thesis}  
Such calculations provide a backscattering amplitude function with a known
(zero) width for absolute comparison to real atom pairs.  Although only 
experience will allow us to determine how accurate
absolute measurements of correlated broadening factors really are, our limited
experience indicates that they are probably within 10\% in most cases.

Another reason direct comparisons are rarely made is simply that diffraction
and XAFS width parameters measure different quantities.  Diffraction relies
on coherent diffracting of x-rays off many crystal lattice planes to generate
a diffraction spot, and therefore its measurements of atom positions and 
broadening factors gives an average over many unit cells.  XAFS depends only
on the instantaneous (within $\sim$10$^{-15}$ sec) position of neighboring
atoms with respect to the central (absorbing) atom, and therefore the broadening
factors are a measure of variations in an atom-pair's distance.  For example,
if two neighboring atoms are vibrating with the same (spatial) amplitude $W$
but are somehow rigidly connected, the bond length would never change, and XAFS
would measure a zero width.  Diffraction would measure broadening factors for
the two atoms to be $W$.  Mathematically, if $\delta r_X$ is the 
instantaneous deviation of atom $X$'s position from its mean position, the
average deviation in the distance between atoms $A$ and $B$ is given by
\begin{eqnarray}
\nonumber
\sigma_{AB} & = &\sqrt{<(\delta r_A-\delta r_B)^2>}\\
& = &\sqrt{<\delta r_A^2>+<\delta r_B^2>-2<\delta r_A \delta r_B>}.
\end{eqnarray}
If we define $\sigma_A=\sqrt{<\delta r_A^2>}$ (which equals $\sqrt{U}$ in
conventional diffraction notation), then the last term can vary between 
$-2\sigma_A \sigma_B$ and $+2\sigma_A \sigma_B$, depending on the 
degree to which the
motions of the two atoms are correlated.  One can define a ``correlation''
parameter $\phi$ from
\begin{equation}
\label{phi_eq}
\sigma_{AB}^2 =\sigma_A^2+\sigma_B^2-2\sigma_A\sigma_B\phi.
\end{equation}
Since $\sigma_A^2$ is given by diffraction
experiments, we can provide
measurements of correlation factors.  $\phi$ will vary between $+1$ for atoms
vibrating in phase with each other (like in an acoustical phonon), $0$ for
atom-pairs that are separated by a few unit cells at temperatures well above
the Debye temperature, and $-1$ for pairs that are moving out-of-phase
with each other (like in an optical phonon).  Of course since the XAFS
measurements give average atom-pair distances over the crystal, static
disorder can produce the same results.

\subsection{Data Reduction}

An initial estimate of $\chi(k)$ is obtained by approximating $\mu_0(E)$ in
Eq. \ref{XAFS_def} by a cubic spline though $\mu(E)$, $k$ by 
$\frac{\sqrt{2m(E-E_0)}}{\hbar}$ and $E_0$ by the energy at the half-height
of the edge.  The data is fit in $r$-space as described in Ref. 
\onlinecite{Hayes82}.  The fit is then subtracted from the data to obtain
a residue which can be utilized to make a better estimate for $\mu_0(E)$, as
described in Ref. \onlinecite{Bridges95b}.

The data have been corrected for the additional ``self-absorption'' of the
fluorescing photon in the sample using a treatment which goes slightly beyond 
the usual correction (for instance, see Tr\"{o}ger {\it et al}\cite{Troger92})
by including the finite thickness of the sample and the effect of the XAFS 
oscillations in the correction term.  The correction for the 5000 \AA{ } film
is small ($\sim$ 8\% of $\mu_0(E)$), but for the 17 $\mu$m single crystal the 
correction is more significant ($\sim$ 60\% of $\mu_0(E)$).

\begin{figure}
\psfig{file=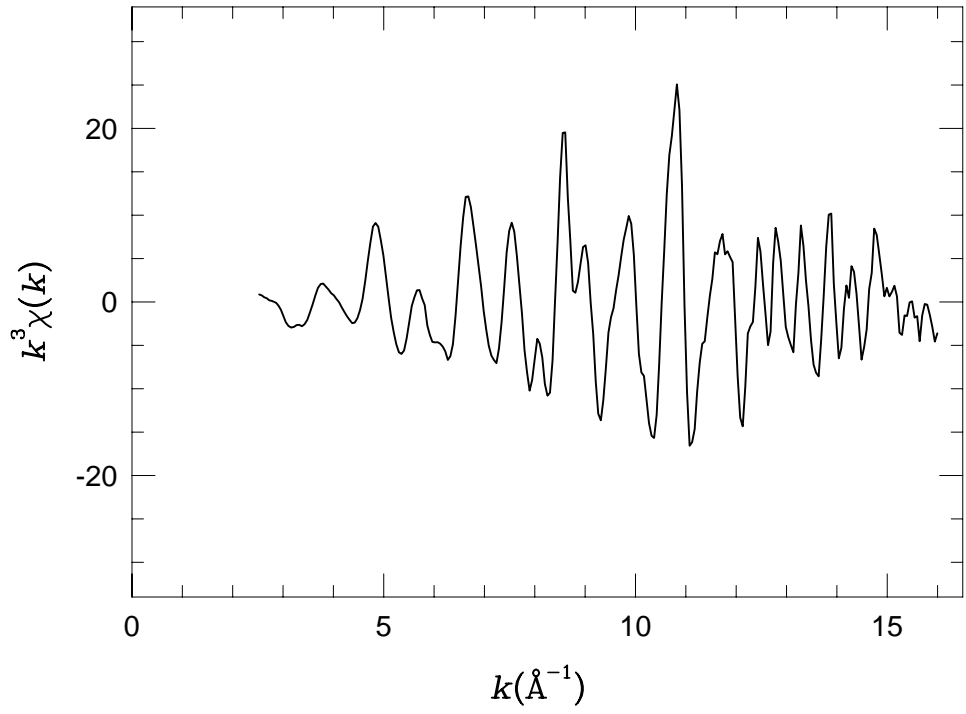,height=8in}
\caption{$k^3\chi(k)$ vs. $k$ for the same sample as in Fig. \protect\ref{e-space}.
}
\label{k-space}
\end{figure}

\begin{figure}
\psfig{file=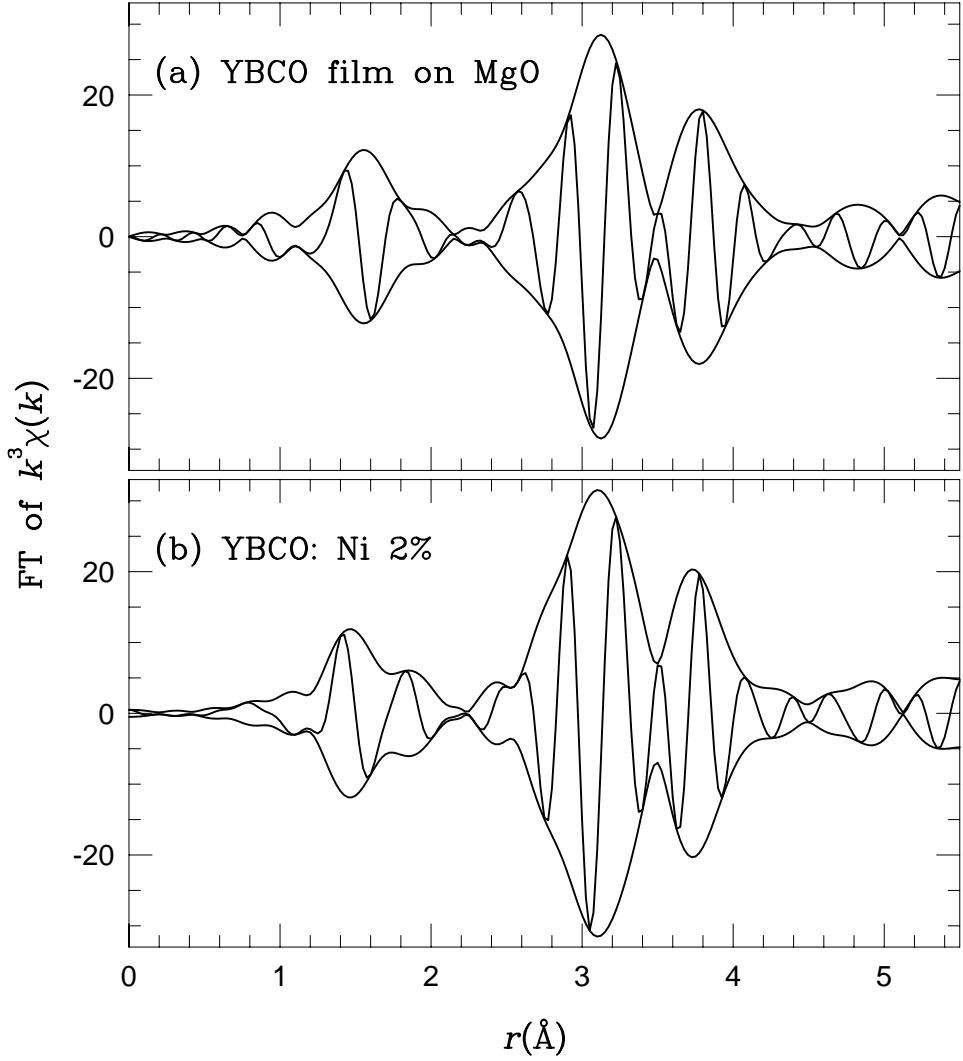,height=8in}
\caption{
The Fourier transform (FT) of $k^3\chi(k)$ ($\equiv \tilde{\chi}^3(r)$) vs.
$r$ for (a) the same sample as in Fig. \protect\ref{e-space}, and (b)
the single crystal of YBCO: Ni 2\%.  Data are transformed 
with a Gaussian window between 3-15.5 \AA$^{-1}$ and broadened by an additional 
0.3 \AA$^{-1}$.  
The oscillating curve is the
$Re(\tilde{\chi}^3)$, and the envelope of this curve is the amplitude,   
$[Im(\tilde{ \chi }^3(r))^2 + Re(\tilde{ \chi }^3(r))^2 ]^\frac{1}{2}$. 
The peak at 1.55 \AA{ } corresponds to the Cu(1)-O(4)
atom pair, while the shoulder of this peak at 2.0 \AA{ } corresponds to the
Cu(2)-O(4) atom pair.  The multi-peak between 2.4 and 3.4 \AA{ } is
a combination of the Cu(2)-Y, Cu(2)-Cu(2), Cu-Ba and Cu(2)-O(2,3) atom
pairs.  Lastly, the peak at 3.8 \AA{ } is the Cu(1)-O(4)-Cu(2) multiple
scattering peak.  The peak positions are shifted from the actual pair-distances
because of the combined effect of $\delta_c$, $\delta_i$ and $F_i(k)$
in Eq. \protect\ref{XAFS_eq}.  
}
\label{r-space}
\end{figure}

An example of $k^3\chi(k)$ for both the film and the single crystal which has
been reduced using the above procedures is displayed in Fig. \ref{k-space}.
The Fourier transform (FT) of $k\chi(k)$ for these data are shown in 
Fig. \ref{r-space}.

\section{Isolating the XAFS signal due to the O(4) contribution}
\label{O4_isolate}

\begin{figure}
\psfig{file=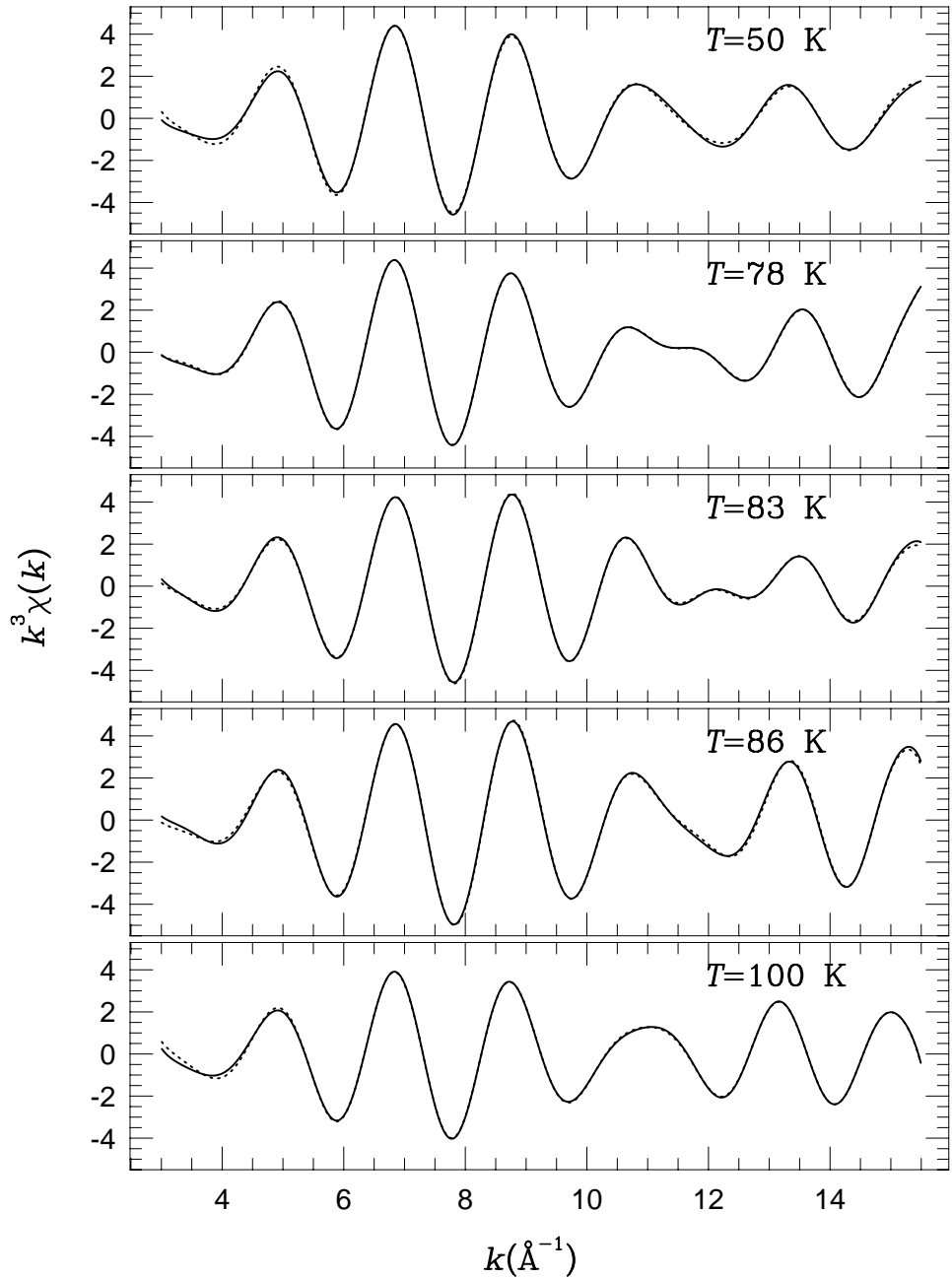,height=8in}
\caption{O(4) contribution to $k^3\chi(k)$ vs. $k$ at several temperatures
for the single crystal of YBCO: Ni 2\%.  The solid lines are the data and the
dotted lines are the fits.  Each fit assumes a split O(4) site except the $T$=100 K 
fit.  The O(4) contribution was obtained in an identical fashion to the data 
in Fig. \protect\ref{O4_film}.
}
\label{O4_xtal}
\end{figure}

\begin{figure}
\psfig{file=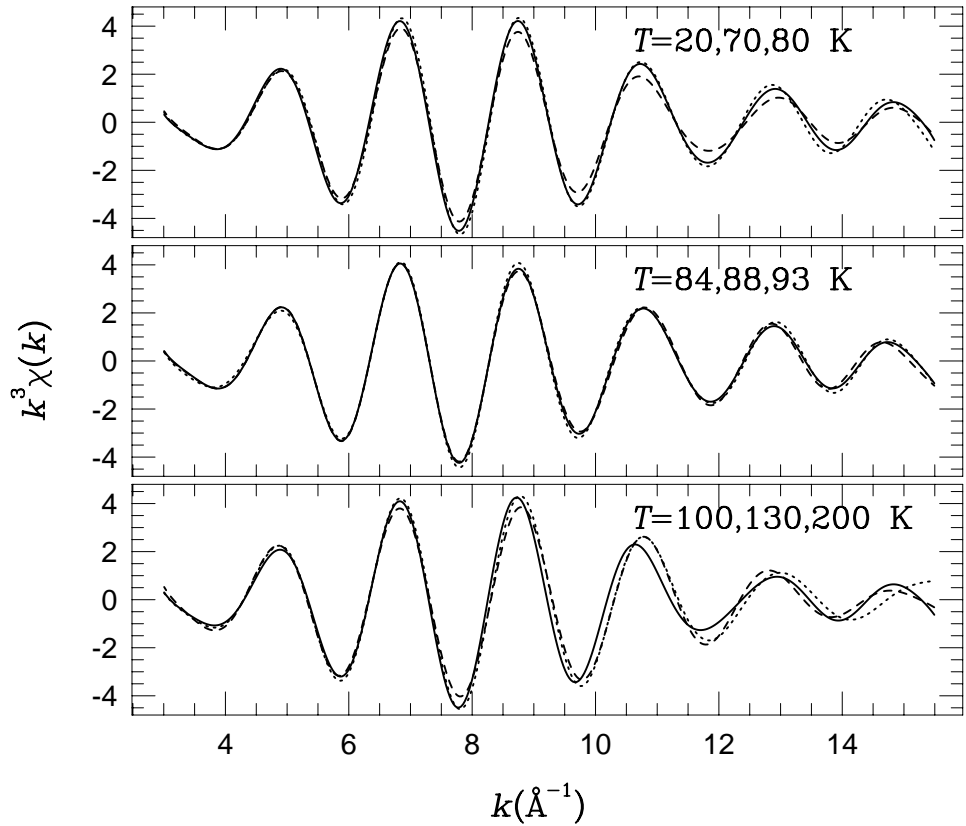,height=8in}
\caption{O(4) contribution to $k^3\chi(k)$ vs. $k$ at several temperatures
for the YBCO film on MgO.  The O(4) contribution was obtained
by back-transforming $r$-space data (Fig. \protect\ref{r-space}) between
1.3-2.2 \AA.  Contributions from high shells (say, from the Cu(2)-Y peak) have
been shown to be negligible.\protect\cite{Leon92}  Only half the data is 
displayed for clarity;  an additional scan was taken in between each of 
the temperatures displayed here.
}
\label{O4_film}
\end{figure}

Since the information above 2.5 \AA{ } is complicated, we
had to perform fits to the data to discern any information about the Y, Ba and
Cu environment.  These fits are described in Sec. \ref{Fits}.  However, since the 
Cu(1)-O(4) and Cu(2)-O(4) peaks are relatively 
well separated, we can isolate them and back-transform the data to $k$-space.
Mustre de Leon {\it et al} based their argument that the O(4) position was split
by observing anharmonic
behavior in the Fourier filtered $k$-space data in the form of a beat 
near 12 \AA$^{-1}$.  
If a two-site distribution exists, we expect the contributions of the two
atom-pairs to be split into four distinct distances:  two corresponding to the
mean Cu(1)-O(4) distance $\pm$ the splitting ($R_{Cu(1)-O(4)} \pm \Delta_r$)
and two corresponding to the mean Cu(2)-O(4) distance $\pm$ the splitting 
($R_{Cu(2)-O(4)} \pm \Delta_r$).  Since the XAFS signal for each of these 
distances is oscillatory, a small splitting will generate a beat in the
XAFS signal at a $k$-vector given by 2$k\Delta_r=\pi$.  In
order to make our data directly comparable to their's, we have also isolated
the O(4) contribution in this way.    First, the FT of
$k^3\chi(k)$ was obtained as in Fig. \ref{r-space}. 
We chose $k^3$-weighting for this transform because the
O(4) peak is much better resolved than in a $k$-weighted transform.  
The Fourier transforms of the $k^3 \chi(k)$ data were then back-transformed from 
1.3 to 2.2 \AA{ }.  The O(4) contribution
from this procedure is displayed for the single crystal in Fig. \ref{O4_xtal} and
for the film on MgO in Fig. \ref{O4_film}.  The sinusoidal behavior in
both these figures is the sum of the Cu(1)-O(4) and the Cu(2)-O(4) components.

The O(4) contribution in the single crystal displayed 
in Fig. \ref{O4_xtal} is in rough agreement with the original work of Mustre de Leon
\cite{Leon90};  although the data in the 11-13 \AA$^{-1}$ range varies from
temperature to temperature, the $T$=83 K and the $T$=86 K data show clearly
anharmonic behavior near 12 \AA$^{-1}$, indicating a splitting of about
$\Delta_r \simeq$ 0.13 \AA.  In fact, except for the $T$=100 K
data, none of the single crystal data can be fit with a single, harmonic
distribution for the O(4) site (Sec. \ref{fit_results}).  

The data from the film on MgO in Fig. \ref{O4_film}, on the 
other hand, displays harmonic behavior at all temperatures and shows very 
little change from temperature to temperature.  This harmonic behavior 
persists out to 16 \AA$^{-1}$.  Since a beat may still occur at higher
wave vectors, these Fourier-filtered $k$-space data set an immediate upper
limit of 0.10 \AA.  The character of the film data is distinctly
different than the single crystal data, showing little if any splitting.
This result also applies to the film on LaAlO$_3$ and therefore appears to be
a generic result of YBCO films.

\section{Fits}
\label{Fits}

Although the raw data has provided much useful information about the nature 
of
the O(4) site distribution, we can still learn more details both about the 
O(4)
distribution and about the further neighbors by performing fits of the 
spectra 
to the sum of theoretical standards.  Such fits provide more quantitative 
details about the local environment around the copper atoms and allow us to
deconvolve the contributions of overlapping peaks due to the Cu(2)-Y,
Cu(2)-Cu(2) (between the planes), Cu(2)-Ba, Cu(1)-Ba and the Cu(1)-Cu(2) pairs.
Furthermore, by fitting the data to theoretical standards calculated by
FEFF6 we can obtain absolute estimates of the broadening factors
$\sigma$ and compare them to results from diffraction studies, thus allowing
for measurements of the correlation parameter $\phi$ (Eq. \ref{phi_eq}).

We only report fit results as a function of temperature for the YBCO film on 
MgO data beyond the near-neighbor Cu(1)-O(4) and the Cu(2)-O(4) atom pairs.

\subsection{Fitting procedures}
\label{fit_proc}

Although weighting the XAFS $\chi(k)$ by a factor of $k^3$ enhances the high $k$
data where the beat in the O(4) part of the spectrum is likely to occur, we have
found that fits to data weighted by a single factor of $k$ give more reliable
results overall.  In general, if we are not focusing on the O(4) part of 
the spectrum, we
fit our data to the FT of $k\chi(k)$.  Fitting to the FT reduces the effect of 
overlapping peaks.\cite{Hayes82}  Since XAFS can be thought of as a sum of the 
contribution to the absorption of the individual atom pairs, we calculate
$F(k)$, $\lambda(k)$, $\delta_c(k)$ and $\delta_i(k)$ given an
approximate atom cluster for each scattering path using FEFF6.
This information for each path is called a ``standard.'' 
The data are fit to a sum of theoretically calculated standards which include
all single scattering paths up to the Cu(1)-Cu(2).  The only multiple scattering
paths that are included involve the Cu(1)-Cu(2) path and the intermediate O(4).
The Cu(2)-planar oxygen path between the planes was too low in amplitude to
obtain reliable fit results, so its parameters were fixed.  Other multiple 
scattering paths were found to be negligible, or were dominanted by the
 single scattering paths.

Several of the parameters were constrained in order to decrease the number of
fit parameters.  The Cu(2)-Y and the Cu(2)-Cu(2) distances were constrained
such that the yttriums rest in the center of the four Cu(2) neighbors given the
diffraction measurements of the $a$ and $b$ lattice parameters of typical
samples.  This constraint was
necessary to fit the Cu(2)-Cu(2) peak (3.38 \AA), which is small compared to the
nearby Cu(2)-Y (3.21 \AA), Cu(2)-Ba (3.37 \AA), and Cu(1)-Ba peak (3.48 \AA).
Also, if the Cu-Ba peaks were allowed to vary over a broad range, they would
often fall into a false minimum in the fitting parameter such that they were 
$\sim$ 0.04 \AA{ } too long.  This behavior was noted by 
Stern {\it et al}.\cite{Stern93}  However, we were able to determine by
fits to a CuI standard (a Cu-I standard should be very similar to a Cu-Ba one)
that this error was due primarily to the overlap
of several neighboring peaks in YBCO, and not to unusual errors in the 
FEFF6-generated
backscattering amplitude.  We therefore only allowed the Cu-Ba peaks to vary 
over a narrow range ($\pm$0.02 \AA{ } from the diffraction result).  All 
width parameters were allowed to vary freely, except $\sigma_{Cu(2)-O(2,3)}$ 
(across the planes), which was held fixed.
$E_0$'s for like-atoms were held equal, but were allowed to vary between 
different  types of atoms to help correct for minor errors in the FEFF6 phase
calculations.  

The aim of these fits is to identify parameters that change with temperature.
Since measurements of $S_{0}^2$ and $\sigma$ are correlated, we 
performed a fit where these parameters were allowed to vary freely, and then
calculated the average for $S_{0}^2$ for each peak.  A similar correlation
occurs between measurements of $R_i$ and $E_0$, so average $E_0$'s for the
individual peaks were also determined.
The values of 
$S_{0}^2$ and $E_0$  were then 
fixed for each path and the fits were performed again.  By
holding $S_{0}^2$ and $E_0$ fixed for all temperatures, we ascribe all changes
in the peaks' amplitude and frequency with temperature to changes in $\sigma_i$
and $R$.

Although 
$S_{0}^2$ should be the same for all paths in principle, we found that 
if a single value of $S_{0}^2$ was used high quality fits could not be obtained 
for the Cu(1)-Cu(2) peak (which includes multiple scattering off the O(4)
atom).  The fit was improved by allowing this one peak to have an 
$S_{0}^2$=0.75, while all other peaks use an $S_{0}^2$=0.90.

The maximum number of 
parameters given our fit range using the method of Stern\cite{Stern93b} is 29.
The total number of parameters in these fits was 17, well below the maximum.  

Errors in the parameter measurements are difficult to estimate reliably in XAFS fits.
We estimate the errors on all the parameters from the covariance matrix
generated by the fit.  The variance of the data used in the covariance matrix
is obtained by assuming that the residual difference between the data and the 
fit is normally distributed.  This
procedure only accounts for random errors and thus should describe relative
errors between data at nearby temperatures.  Systematic errors due to problems
in the FEFF6 calculation, self-absorption corrections, estimates of $\mu_0$,
etc., will cause overall shifts in the best fit.  
Absolute errors on these measurements are generally 
$\lesssim \pm 0.02$ \AA{ } in $R$ and $\lesssim \pm 15\%$ in $S_{0}^2$ 
and $\sigma$, and are generally better for nearest neighbor.\cite{Li95b}

\subsection{Fitting results}
\label{fit_results}

\subsubsection{O(4)-site distribution in the single crystal}
\label{fit_results_xtal}

As suggested by the presence of the beats, a two O(4) site distribution
was necessary to fit the single crystal data in Fig. \ref{O4_xtal} for the
four lowest temperatures.  The fits shown in 
Fig. \ref{O4_xtal} all assume a two-site distribution for the O(4) except for
the $T$=100 K data.  For the $T$=100 K data, a single site for the O(4)
produces a reasonable fit; the slight dip in the amplitude at 
$\sim$ 11 \AA$^{-1}$ can be modeled by an interference between a single
Cu(1)-O(4) path and a single Cu(2)-O(4) path.  Note that the data in
this region is still visibly different from the film data.
The features in the $T$=50, 
78, 83, and 86 K data between 10 and 13 \AA$^{-1}$ are modeled by two 
Cu(1)-O(4) paths and two Cu(2)-O(4) paths, with each path length split by 
0.12 $\pm$ 0.01 \AA{ } and 0.06 $\pm$ 0.06 \AA, respectively.  We did not 
observe any obvious trend with temperature other than for the $T$=100 K data.

The two-site fits to the single crystal data may be deceptively good.  In
order to obtain these fits, unphysically small ($<$ 0.01 \AA)  values for the 
broadening factors were needed.  In addition, the sum of the amplitudes for these 
peaks (which should add up to the number of neighbors times $S_0^2$) is about 
50\% too high.  These exceptionally narrow peaks indicate that besides the split
peak distribution, further anharmonicity must exist. A common way to model 
anharmonicity in XAFS data analysis is to expand Eq. \ref{XAFS_eq}
about $R_i$ for a given atom-pair in powers of $<r^n>$, otherwise known as a 
``cumulant expansion'' ($<r^n>$ is the $n^{th}$ moment,$M_4$ is the part of
the 4th moment that is different from a gaussian, 
namely $M_4=<r^n>-3\sigma^4$).\cite{Eisenberger79}  
The 2nd cumulant is the width in the harmonic approximation, or 
the Debye-Waller factor.  The 3$^{rd}$ cumulant affects the phase shift.  The 
4$^{th}$ cumulant multiplies the overall amplitude by a factor 
$e^{\frac{2}{3}k^4 M_4}$.  If this term is positive, it {\it increase} the 
amplitude of the XAFS 
at higher wave vectors, acting like an imaginary Debye-Waller factor, or
like a distribution with a cusp which puts more weight at the center of the
distribution at the expense of the wings.\cite{NR_mom} 
Excellent fits to the single crystal data can be obtained by allowing the four 
Cu-O(4) peaks to have $M_4\cong2.7\times10^{-5}$ \AA${^4}$.  These fits have 
much more reasonable Debye-Waller factors ($\sim$0.04 \AA) and $S_0^2$ is no 
longer 50\% too large.  With such a large value of $M_4$, the cumulant expansion
is no longer accurate above a $k$ of $\sim 14$ \AA.

These fits serve to illustrate that we can model the O(4) site distribution 
with reasonable values for the fit parameters, but that the fits are 
not unique.  Because $S^2_0$, $\sigma$ and $<r^4>$ are so correlated 
for these slightly separated Cu-O(4) peaks, a study of the temperature
dependence of any of these parameters would produce questionable
results.
The only firm result is that at least three, and possibly four,
 atom-pair distances are 
necessary to describe the beat structure in the $k$-space transforms
for the single crystal.  The actual
O(4)-site distribution $g(r)$ may be more complicated.  

Table \ref{thetable} reports fit results to the further neighbors for the $T$=50 K
sample.

\subsubsection{Fits to the film on MgO}

\begin{figure}
\psfig{file=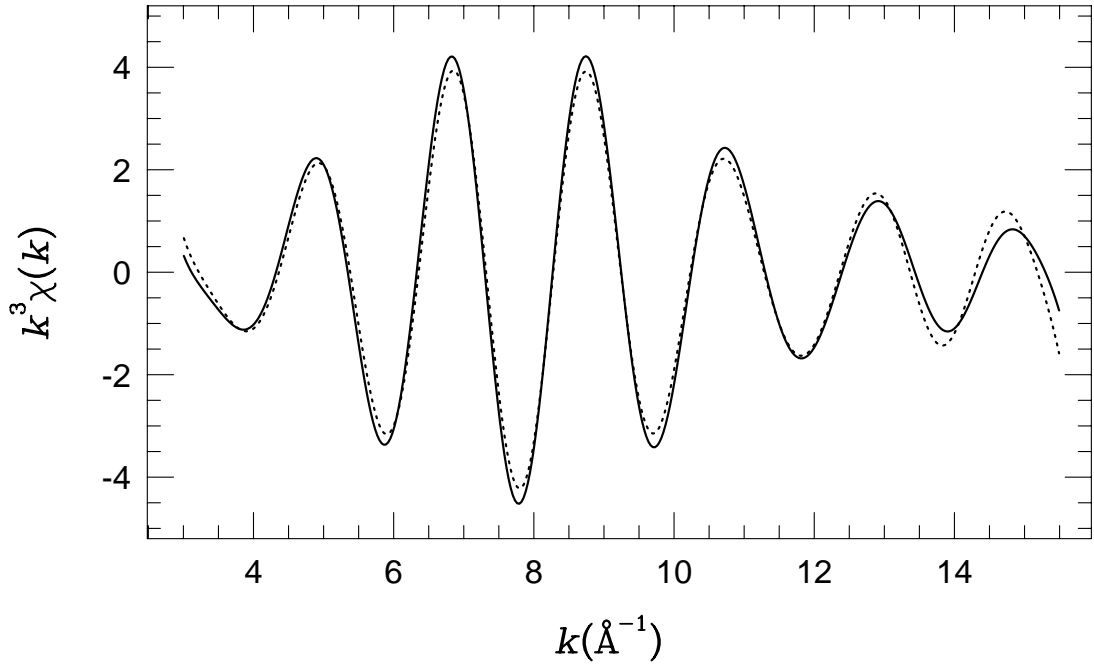,width=7in}
\caption{Fit to the O(4) contribution to $k^3\chi(k)$ vs. $k$ for
the YBCO film on MgO at $T$=20 K.  Fit assumes a single O(4) site.
}
\label{O4_film_fit}
\end{figure}

\begin{figure}
\psfig{file=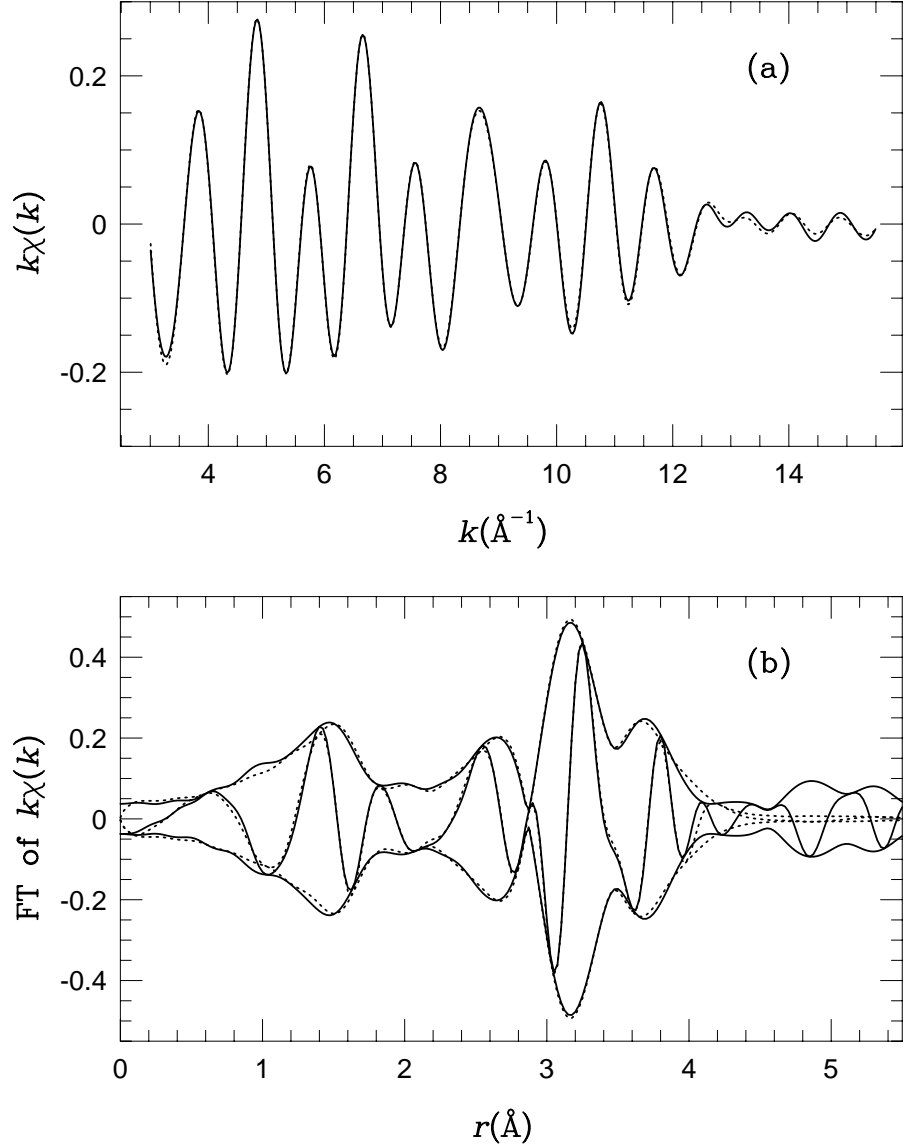,height=8in}
\caption{Full fit to (a) $k\chi(k)$ vs. $k$ and (b) FT of
$k\chi(k)$ vs. $r$ for the YBCO film on MgO.  Data in (a) are the
back transform of the the data in (b) from 1.3-4.0 \AA{ } in $r$-space.
FT range in (b) is same as in Fig. \protect\ref{r-space}.  
Notice that the quality of the
fit is very good at low $k$ and degraded somewhat at high $k$.
}
\label{full_film_fit}
\end{figure}

The Cu(1)-O(4) and Cu(2)-O(4) peaks are well described by a single O(4) site
in the film data, as suggested in Sec. \ref{O4_isolate} 
(Fig. \ref{O4_film_fit}).  
Allowing for a two site distribution with a $\Delta_r \gtrsim $ 0.09 \AA{ }
did not produce reasonable fits.  
Fits to the film data of a similar quality as to the single crystal data 
are obtained without assuming any anharmonic
behavior in the atom-pair distances out to $\sim$ 4.1 \AA.  
Fig. \ref{full_film_fit} shows the fit to the film data over this 
range both in $k$-space and in $r$-space, and Table \ref{thetable}
reports the fit to all the parameters for both the single crystal
and the film on MgO at $T$=50 K.  No significant
deviation between the XAFS atom-pair distance measurements and standard
diffraction measurements were observed, except for the unusually short 
Cu(2)-O(4) and Cu(1)-Cu(2) pairs, which are about 0.02 \AA{ } shorter than in 
the single crystal diffraction measurements.  This measurement is consistent 
with the shorter $c$-axis of this material (11.64 \AA{ } at room temperature) 
compared to the diffraction sample (11.68 \AA{ } at room temperature).  

\begin{figure}
\psfig{file=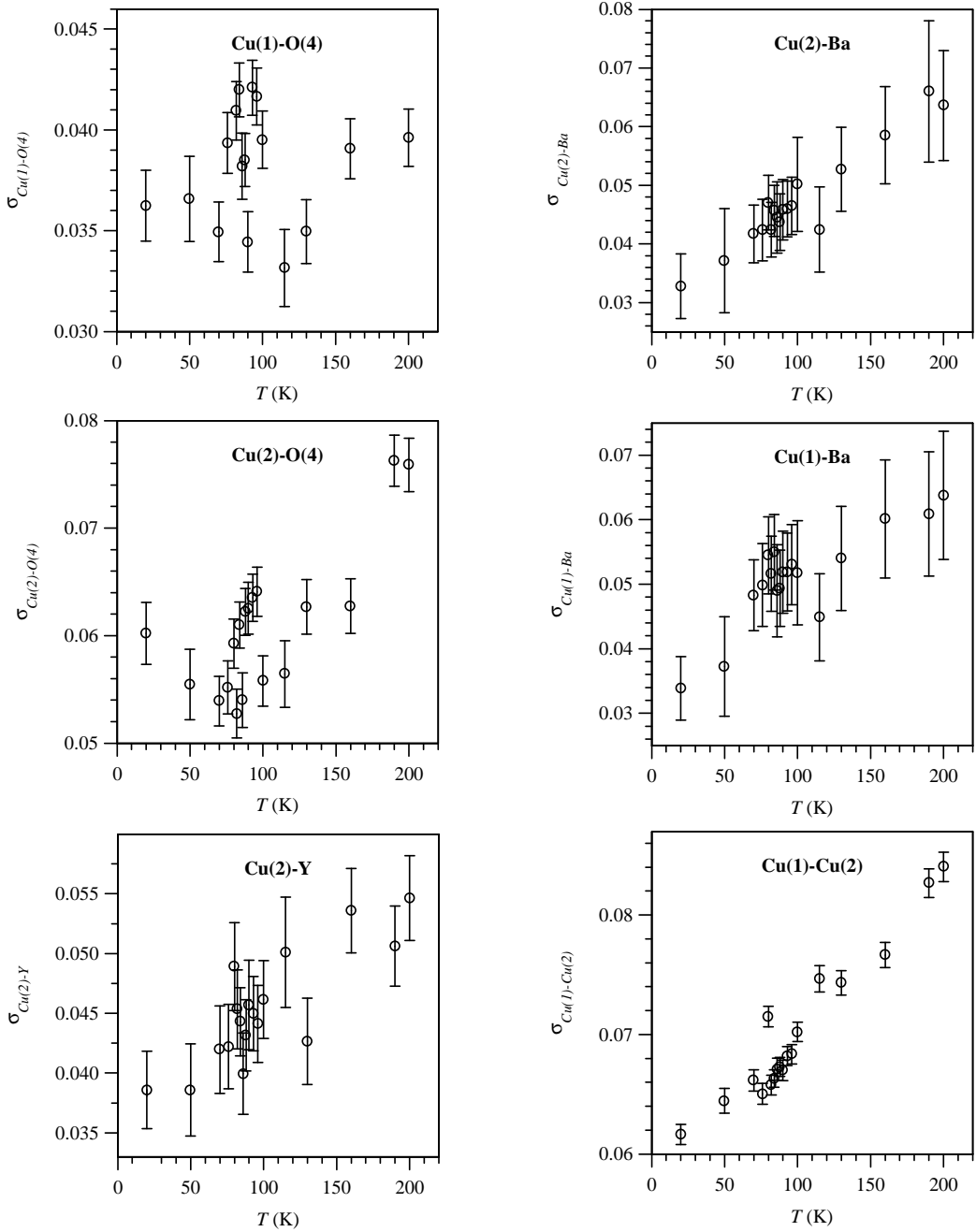,height=9in,bblly=220pt,clip=}
\vspace{-1in}
\caption{ $\sigma$ vs. temperature from the film of YBCO on MgO for all 
single scattering paths up to
4.15 \AA, except the Cu(2)-Cu(2) (at $\sim$3.38 \AA) and the Cu(2)-O(2,3) 
(at $\sim$3.66 \AA) pairs.  Errors are estimated from the 
covariance matrix, and in no way reflect any systematic errors, as discussed in 
Sec. \protect\ref{fit_proc}.  These errors should thus be taken as relative errors which indicate the
reproducability of the data.
}
\label{sig_fig}
\end{figure}

Plots of $\sigma$ vs. $T$ are shown in Fig. \ref{sig_fig}.  All the atom-pairs 
show an increase of $\sigma$ with temperature, except the Cu(1)-O(4)  and the
Cu(2)-Cu(2) (not shown) atom pairs.  The measurements of $\sigma_{Cu(2)-Cu(2)}$ 
are fairly
large ($\sim$ 0.11 \AA) with large estimated errors and are therefore unreliable.
Only the Cu(1)-O(4) and the Cu(2)-O(4) pairs show any anomalies in $\sigma$ near 
$T_c$.  $\sigma$ for both pairs increases slightly just above $T_c$ and then 
returns just below $T_c$.  The jump is larger for the Cu(2)-O(4) width 
($\sim$0.008 \AA) than for the Cu(1)-O(4) width ($\sim$ 0.004 \AA).  This 
behavior was noted 
for the Cu(1)-O(4) bond both by Stern {\it et al}\cite{Stern93} and in Kimura {\it et al}\cite{Kimura93}, although the Stern result assumed a split O(4) site.
None of the previous XAFS studies report a change in the Cu(2)-O(4) width.

All other atom-pair parameters show smooth behavior with temperature and are 
of reasonable values.

\subsection{Correlations of the further neighbors}

We have analyzed the 
correlations between positions of the Cu atoms and their near-neighbors by
calculating $\phi$ for each atom pair.  
Diffraction measurements of broadening factors for each site are required,
as indicated in Eq. \ref{phi_eq}.
We have chosen temperature-dependent neutron diffraction data given by 
Sharma {\it et al}\cite{Sharma91} because they include a relatively 
dense grid in temperature between data points ($\Delta T\cong$10 K), 
measurements over a similar range of temperatures (10-300 K) 
to this study, anisotropic thermal factors for the O(4) site, and a high 
sensitivity to the oxygen atoms.  This diffraction
study saw no anomalous features with temperature in the thermal factors for any 
lattice site and is consistent with a similar study by 
Kwei {\it et al}.\cite{Kwei90} Since
Sharma {\it et al}'s measurements of the thermal factors were not always at the 
same temperatures as the measurements in this report, we fit each Debye-Waller 
factor from diffraction vs. temperature with a polynomial.  The use
of these thermal factors may introduce systematic errors into our measurements 
of $\phi$
not only because the samples are prepared with different methods in different 
laboratories, but also because the diffraction
results were obtained from a powder of YBCO and our data is for a film.
Unfortunately, we must tolerate such errors because no comprehensive, 
temperature-dependent study of a thin film of YBCO exists at this time.  Such 
errors should only contribute to an overall shift of the $\phi$ parameter and 
should not greatly affect the temperature dependence unless major features in 
the (as yet, unmeasured) diffraction thermal factors occur in measurements
on films and not on single crystals.

\begin{figure}
\psfig{file=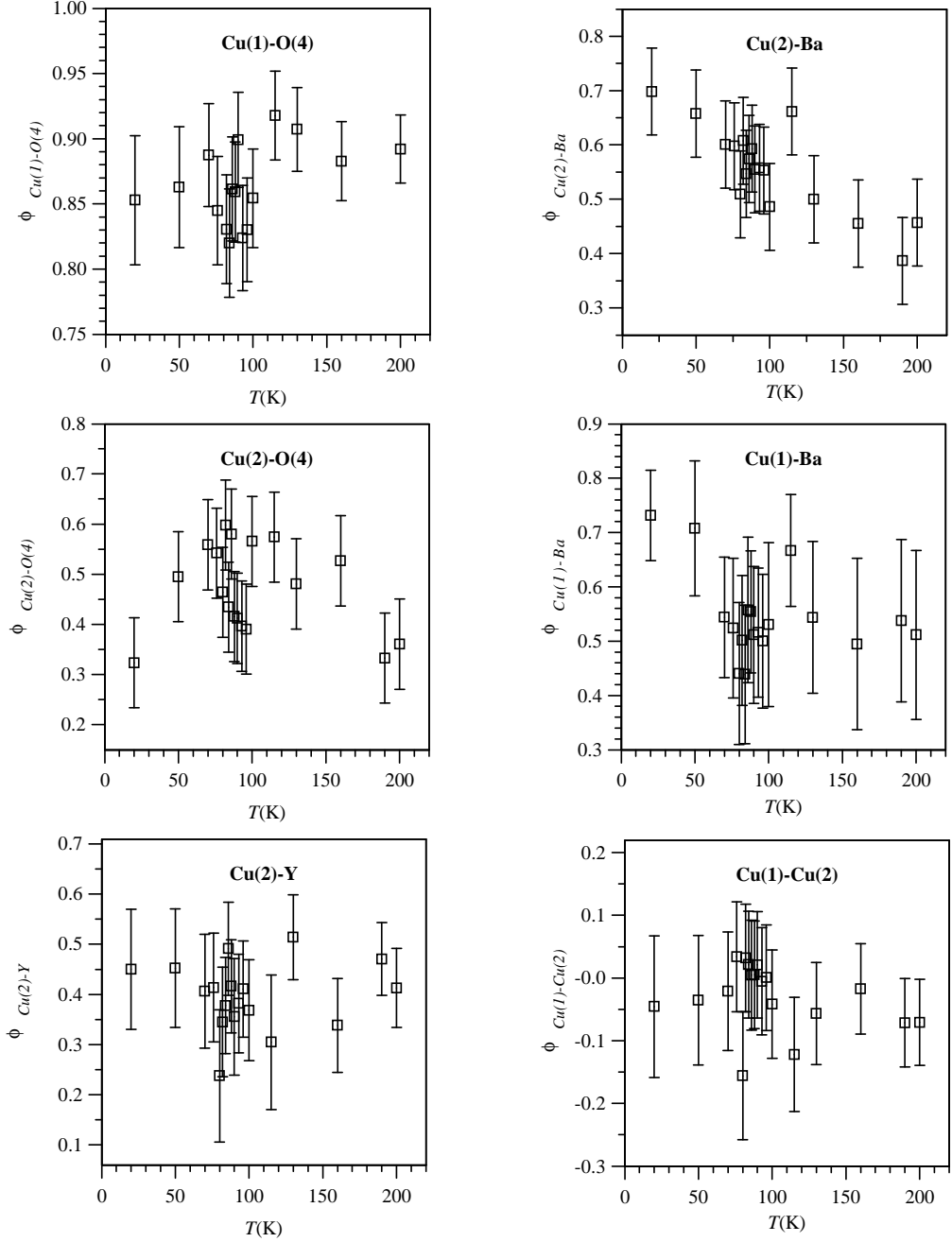,height=9in,bblly=220pt,clip=}
\vspace{-1in}
\caption{The correlation coefficient $\phi$ vs. temperature from the film of 
YBCO on MgO for the same atom pairs as in Fig. \protect\ref{sig_fig}.  Errors
are propagated through Eq. \protect\ref{phi_eq} using the errors for the diffraction data 
reported in Ref. \protect\onlinecite{Sharma91}.
}
\label{phi_fig}
\end{figure}

The correlation coefficients vs. temperature for each atom pair are displayed 
in Fig. \ref{phi_fig}.  Errors
are propagated from the quoted errors in Ref. \onlinecite{Sharma91} and the
errors in Fig. \ref{sig_fig} and in no way attempt to reflect any systematic
errors introduced by FEFF6, differences in the samples, etc.  The correlations
for the Cu(2)-Cu(2) bond are not displayed because our measurements
of $\sigma_{Cu(2)-Cu(2)}$ are unreliably large with large estimated errors.
All the other atom-pairs are resolved in the fits well enough to give 
reasonable estimates of $\phi$.  

\section{Discussion}
\label{Discussion}

\subsection{Split O(4) site}

The main result reported above is that while two O(4) sites are necessary
to describe the
XAFS data for a single crystal very well, a high-quality fit
to XAFS data for the thin film with a single O(4) site is also possible.  
Although this result only
puts an upper limit on the size of any site splitting at $\lesssim$0.09 \AA,
the measurements of the Debye-Waller factors are small enough to indicate
that any possible splitting is probably even smaller.  (Any unresolved
splitting should add to the Debye-Waller factors in quadrature.)  

Even though the existence of the splitting in some samples is now well established,
the nature of the splitting is still quite unclear.  The main problem
is that the split sites could be due to motions of individual atoms between distinct
sites, or
due to two separate harmonic potentials, displaced from each other by, perhaps,
a local distortion.
One model\cite{Rohler94} for static splitting relies
on oxygen vacancy ordering on the O(1) site, which may shift the local position
of the O(4) atom.  At least one configuration of oxygen vacancies can produce
a split O(4) position when half of the O(1) sites are occupied:  
the Ortho-II phase of YBa$_2$Cu$_3$O$_{6.5}$, which occurs
when every other Cu-O(1) chain is completely vacant of oxygen.  In this case,
only half the O(4)'s have near O(1) neighbors.  This
configuration splits the O(4) site into two sites separated by $\pm$0.05 \AA{ } from the
stoichiometric (no O(1) vacancies) site.\cite{Zeiske91,Zeiske92b}  
R\"{o}hler\cite{Rohler94} considers a similar situation when single O(1)
vacancies cause shifts in the O(4) position both along the same chain and
along neighboring chains.  In this model a single O(1) vacancy 
affects 12 O(4)
atoms.  In the case of optimally doped YBCO with $\delta=6.93$, this would
mean 84\% of the O(4) oxygens could be affected.  This model also helps
explain why the effect seems to be more easily fit as a split O(4) 
distribution in lower $T_c$ samples
(i.e. more O(1) vacancies) as measured by Stern {\it et al}.\cite{Stern93}  Any changes
observed in the splitting near $T_c$ would be considered a reordering of oxygen 
vacancies in this model.  In order to explain the thin film results, one
could invoke the added stress on the material introduced by the substrate.
This stress could perhaps be manifest as a freezing of the vacancies 
into a random organization
that suppresses the amount of splitting.  This hypothesis is not
likely, however, because oxygen diffusion measurements are not drastically
different in a film compared to crystals.\cite{Lee92,Pashmakov95}  Another possibility is
that the films are overdoped and hence there are very few O(1) vacancies.  A
direct test of this argument would be to measure films that range from overdoped to
underdoped. 

A dynamic model that would lead to either a broadening or a split O(4) 
position is the hopping small polaron model, with the polaron hopping 
on and off the Cu-O(4) bonds as discussed by 
Mustre de Leon {\it et al.}\cite{Leon92}  Ranninger\cite{Ranninger92} has calculated
the XAFS signatures for the presence of polarons and shown that as the 
hopping speed decreases, the pair distribution function first broadens 
and then splits;  less than a factor of two in hopping speed is required
to change from a broadened peak to a split peak.  This is expected from 
the following simple argument.  When the electron hops at a much faster rate
than the lightest optical phonons, the lattice cannot respond, and there
is little displacement of the individual atoms.  If the hopping is slower,
the lattice has time to respond, and the Cu and O(4) atoms move towards
each other or apart as the polaron hops on and off the ligand.  If the
hopping rate falls below the optical frequency, a split Cu-O(4) distribution 
emerges, corresponding to the absence or presence of the polaron.
This model can easily explain both the single crystal and the thin film
results with a small change in the polaron hopping rate.  The fact that the 
Cu(2)-O(4) peak is less ordered than the Cu(1)-O(4) peak and has a 
correlation parameter $\sim$0.5 is consistent with such a model.

Although the oxygen vacancy argument has possible merit, there is 
a body of literature
that links dynamics of the O(4) site to anomalies in vibrational 
spectroscopies both in YBCO\cite{Mihailovic93,Mihailovic91,Nickel93}
and other material.\cite{Mihailovic93,Mihailovic91,Lee94,Booth95}
Since XAFS is
not sensitive to static verses dynamic changes in atomic positions, there is
little chance the XAFS technique can resolve this issue.  However,
XAFS can determine whether a pair-distribution function is harmonic or not.
In our fits to the split distribution in the single crystal, we had to allow
for a large fourth cumulant for both O(4) sites.  This term can either
have the effect of flattening out or sharpening the
peak of an otherwise harmonic distribution, depending on the sign.  These
data required a positive 4th-cumulant, and were thus sharper than a harmonic
distribution.  No third cumulant was necessary to
fit the data.  The necessity of this extra parameter indicates that the 
pair-distribution function is not simply the sum of two harmonic distributions.
The need for this quartic term is
consistent with Raman measurements of the O(4) mode at 505 cm$^{-1}$ which also
requires a significant 4th-order term.\cite{Mihailovic93}  
This anharmonicity has been speculated\cite{Mihailovic93} to be related to the 
anharmonic potential implied by the split O(4).\cite{Leon92,Leon93}
The anharmonicity in the O(4) 505 cm$^{-1}$ mode could now be directly
linked to the split O(4) distribution if accurate measurements of the 
Raman anharmonicity
can be shown to be significantly different between a film and a single crystal.  
This result would mean that the anharmonicity (and any polaron formation related 
to this anharmonicity) is not required for high-$T_c$ superconductivity.  On
the other hand, if the anharmonicity in the 505 cm$^{-1}$ peak is unchanged 
between the film and the crystal, then the split O(4) is not related to it.  In 
this scenario, polaron formation may still exist and be important for high-$T_c$,
but the split O(4) would not be the structural manifestation of it. 

When trying to explain why splitting of the O(4) site may or may not 
be present, one should also consider the $T_c$ dependence.  Stern {\it et al}
reports that the changes in the splitting near $T_c$ are more pronounced for 
the lower $T_c$ samples.  However, the two-site fits to the O(4) for the higher 
$T_c$ samples in their work are not of the same high quality as for the lower 
$T_c$ samples.  In other words, the pair-distribution function for these samples
appears to be more complicated than a simple two-site distribution.
Therefore, the Stern fits cannot really rule out any significant temperature 
dependence.  Certainly our own
results on the single crystal indicate that the temperature dependence may
not be consistent from sample to sample; all the data at $T < T_c$ cannot be 
fit satisfactorily with a single O(4) site, yet the $T$=100 K data is well 
described with a single O(4) site.  This result is in contrast to the work by Mustre de Leon {\it et al}, where they report the splitting to be constant at
all temperatures except near $T_c$ where the splitting shrinks from
0.13 \AA{ } to $<$ 0.11 \AA, and to the work by Stern {\it et al},
where for some samples, they report a constant splitting at all temperatures.  

\subsection{Changes in broadening parameters}
\label{dis_sigma}

In our measurements of the broadening parameters of the film on MgO we observe some
significant temperature dependencies.  Almost all the measured bonds show the 
Debye-Waller factor increasing with temperature, as one expects if the Debye 
temperature is not too large (Fig. \ref{sig_fig}).  The exception is the Cu(1)-O(4)
bond which maintains a broadening of approximately 0.035-0.40 \AA{ } from 
$T$=20 to 200 K.
The only significant behavior occurring near $T_c$ involves the O(4) atom:  both the
Cu(1)-O(4) and the Cu(2)-O(4) pairs show an increase in their broadening parameter,
with $\sigma_{Cu(1)-O(4)}$ jumping from 0.035 to 0.040 between 80 K and 100 K
and back then to 0.035 \AA, and $\sigma_{Cu(2)-O(4)}$ jumping from 0.060 \AA{ }
to 0.068 \AA{ } between 80 K and 96 K and back to 
0.060 \AA.  In the
region between $T$=80 K and 100 K, both $\sigma_{Cu(1)-O(4)}$ and $\sigma_{Cu(2)-O(4)}$
 varies from temperature to
temperature more than the estimated error.  This fluctuation may be real 
or the error estimates may be too small for some reason, perhaps from inaccurate
temperature measurements.  However, the Cu(2)-O(4) data vary more smoothly. 
Stern {\it et al} 
also measured a fluctuation  in the Cu(1)-O(4) broadening near $T_c$, however 
their measurement assumed a two site distribution
that had collapsed to one site in the same temperature range as the 
broadening fluctuation.
Kimura {\it et al}\cite{Kimura93} measured a fluctuation in $\sigma_{Cu(1)-O(4)}$ in a similar 
temperature range using a single O(4) site fit, but reported no anomalies in 
$\sigma_{Cu(2)-O(4)}$
near $T_c$.  No proven explanation of the physics behind these fluctuations exists, 
however, they can be interpreted in terms of changes in the correlation between the 
Cu and O atomic positions (see next section).

\subsection{Correlations between atomic positions}
\label{dis_cor}

Like the O(4)-site position, changes in $\phi$ with temperature can be 
attributed to changes in the static disorder of the individual sites,
or to changes in the dynamics of the lattice displacements.  As the 
temperature of the sample is decreased, static disorder could manifest itself
as an ordering of oxygen vacancies, which would then generate a distinct
set of positions for a given atom.  If the site is split into
two or three distinct positions, XAFS would measure an unusually broad peak for
that pair and thus look like the atom-pair displacements are negatively correlated.
It is therefore important to consider
the absolute value of the broadening measurements from XAFS as compared to diffraction
to try to help determine if the model used to fit the data is consistent, i.e.
single atomic sites broadened harmonically around some average pair distance.
Also, trends in the correlation with temperature should give some insight
into which correlations within the unit cell are dependent on
static or thermal disorder.

The most important result from the measurements of $\phi$ in Fig. \ref{phi_fig}
is that, within the estimated errors, they are all in the range from 0 to 1, 
indicating that the measurements
of $S_0^2$, the calculated $F(k)$ and the model used to fit the data
are all fairly accurate.  More direct tests of the reliability of the absolute
measurements of $\phi$ are provided by the nearest and the furthest bonds 
measured, that is, the Cu(1)-O(4) pairs and the Cu(1)-Cu(2) multiple 
scattering pair.
Since the Cu(1)-O(4) pair is the nearest neighbor pair in YBCO, we expect at 
low temperatures that $\phi$ should be very near unity, and it is:  
$\phi_{Cu(1)-O(4)}$ demonstrates weak temperature dependence 
with a mean value of about 0.87, in approximate
agreement with correlation measurements of the Hg-O(2) pair in 
Hg-1201.\cite{Booth95} The fluctuations near $T_c$ are well within the 
increased errors after propagating the diffraction errors into $\phi$.  

The Cu(1)-Cu(2) atom pair is the furthest neighbor fit and includes multiple
scattering off the intervening O(4) atom. It is therefore a good test of the
FEFF multiple scattering calculations for $F(k)$\cite{FEFF5} as
well as the absolute reliability of $\phi$.  We expect this pair
to be the least correlated,
and it is;  $\phi_{Cu(1)-Cu(2)}$ shows no obvious change with temperature
(although both XAFS and diffraction broadening factors show a significant temperature
dependence)
and is consistent with $\phi_{Cu(1)-Cu(2)}\cong$0.  The absolute accuracy of
this pair is worse than the other pairs measured because the FEFF calculation
includes multiple scattering off the O(4).  Consequently, this measurement
is consistent with $\phi_{Cu(1)-Cu(2)}\cong\pm$0.2.  
The Cu-Ba pairs show a strong temperature 
dependence in $\phi$, in each case dropping from about $\phi$=0.7 to 0.5 between 20 and 
200 K.  
The Cu(2)-Y pair is partially correlated ($\sim$ 0.45 $\pm$ 0.1) and shows little
temperature dependence between 20-200 K. 
The lower correlation coefficient for the cold temperature Cu(2)-Y
pair compared to the Cu-Ba pairs is probably because both atoms in the pair
are bound to the planar oxygens, with bond angles between the Cu(2)-O(2,3)
and Y-O(2,3) bonds $\sim$90$^\circ$;  the Ba atoms are more constrained by the
axial oxygens.  

The most interesting correlation measurements are for the 
Cu(2)-O(4)
pair.  $\phi_{Cu(2)-O(4)}$ starts out very low ($\sim$0.4) but starts to
{\it increase} to a maximum of $\sim$0.6 just below $T_c$.  In the
vicinity of $T_c$, $\phi$ then drops
to about 0.45 and then increases again to its maximum value of 0.6.  As the
temperature is increased above 100 K, $\phi$ decreases steadily from 0.6 
to 0.4.  This decrease in $\phi$ (and hence the increase in $\sigma$ mentioned
in Sec. \ref{dis_sigma}) may be the result of a negatively-correlated 
mode being excited near $T_c$ relative to the correlated mode that is dominating
away from $T_c$.  Such a mode may be consistent with a polaronic-hopping
transport mode.\cite{Booth95}

\section{Conclusions}
\label{Conclusions}

We have contrasted XAFS measurements on a thin film of YBCO on MgO with
measurements on a single crystal with 2\% Ni. We have also compared 
the results with
a film on LaAlO$_3$, which gives essentially the same results, and 
with previous measurements on oriented powders.  
The single crystal data exhibit an
anharmonic site distribution for the O(4) atom which can be described as a
split position below $T_c$ and a single position above $T_c$.  (The 
temperature
range measured for the single crystal is limited.)  This result is consistent
with previous results on oriented powders.\cite{Leon90,Stern93}  

The main result of this paper is that the
thin film data show very different behavior.  The O(4) peak can be well 
described by a single-site, harmonic distribution.  This distribution does
not show any strongly anharmonic behavior with temperature near $T_c$, or at any other
temperature between 20-200 K.

These data refute the argument that the XAFS signal showing the split
O(4) site distribution is too small to be measured reliably;\cite{Stern93}
however, high S/N is required.
The Fourier-filtered film data are very reproducible over the temperature
range measured, and since only subtle changes occur in the O(4) site distribution
for the film near $T_c$, the changes in the data with temperature are gradual and predominantly
thermally driven.  Because of this reproducibility, the rapidly changing XAFS
from the single crystal must be taken as a legitimate signal, and not as noise.

Although we don't see any evidence of anharmonicity in the O(4) 
site in the film, fits show that the broadening factor for 
the Cu(2)-O(4) bond exhibits a local maximum above $T_c$ followed by a drop
below $T_c$ to essentially its $T$=20 K value.  The Cu(1)-O(4) pair demonstrates
similar behavior, but the effect is smaller and is not as systematic with temperature.
Since there is no such behavior 
in the published diffraction literature\cite{Sharma91,Kwei90} we have interpreted
this as a change in the degree of correlation between the Cu(2) and the
O(4) atoms, and possibly between the Cu(1) and the O(4) atoms.  This result
is in contrast to measurements by Kimura {\it et al}\cite{Kimura93}
which show changes in the broadening factor for the Cu(1)-O(4) bond near
$T_c$, but not in the Cu(2)-O(4) bond.  

Finally, a better understanding of the local structure is obtained by considering the
local correlations of the near-neighbors.  As mentioned above, the Cu(1)-O(4)
is measured to be a very tight bond, yet the Cu(2)-O(4) is ``looser" and 
exhibits a decrease in the correlation between the atoms near $T_c$, indicating that the 
chains act as a unit
that is somewhat independent of the planes.  This lack of correlation is
also indicated by the measurement of the correlation coefficient 
$\phi_{Cu(1)-Cu(2)}$ near zero ($<$ 0.2) that also does not change with
temperature.  However, the atomic displacements in the metal-oxide layer do
become less correlated at higher temperatures.  
The atomic positions in the CuO$_2$ planes apparently remain 
correlated in the temperature range measure based on the measurements on
the Cu(2)-Y atom pair.

\acknowledgements

We wish to thank G. G. Li, E. D. Bauer, and Z. Kvitky for many useful
discussions and for proof reading the manuscript.  The experiments were
performed at the Stanford Synchrotron Radiation
Laboratory, which is operated by the U.S. Department of Energy, Division
of Chemical Sciences, and by the NIH, Biomedical Resource Technology Program,
Division of Research Resources.  The experiment is partially carried out on
UC/National Laboratories PRT beam time.
The work is supported in part by NSF grant DMR-92-05204.

\bibliography{/exafs/bib/bibli}
\bibliographystyle{/exafs/bib/prsty}

\onecolumn
\begin{table}[b]
\widetext
\caption{Comparison between fit results to the single crystal of 
YBCO:Ni 2\% (xtal) and the film of YBCO on MgO at $T$=50 K.  {\it n.b.}
stands for "number of bonds per unit cell."  Neutron
diffraction results of Sharm {\it et al}\protect\cite{Sharma91} are
also given for comparison, and measurements of $\phi$ are reported for
the film data.  Measurements of $\phi$ for the single crystal are 
qualitatively the same.  Errors marked with an ``x'' are unreliable
because the pair-distribution function deviates significantly from
a gaussian.  Imaginary Debye-Waller factors often accompany these distributions;
such parameters cause the XAFS oscillations to {\it increase} 
with $k$ (unphysical).  Some parameters were held fixed or constrained in the
fits.  See Sec. \protect\ref{fit_results} for further discussion. Direct
comparisons of bond lengths should consider that at room temperature,
the diffraction sample has a $c$-axis lattice parameter of 11.68 \AA, while
the film's is 11.64 \AA.  The lattice parameters of the single crystal
are not known. }
\begin{tabular}{lllllllllllllll}
&\multicolumn{3}{c}{XAFS xtal}&&
\multicolumn{3}{c}{XAFS film}&&
\multicolumn{4}{c}{Neutron Diff.\protect\cite{Sharma91}}&&film\\
Bond & $r$(\AA) & {\it n.b.} & $\sigma$(\AA) && 
       $r$(\AA) & {\it n.b.} & $\sigma$(\AA) &&
       $r$(\AA) & {\it n.b.} & $\sigma_A$(\AA) & $\sigma_B$(\AA) &&
       $\phi$\\
\tableline
Cu(1)-O(4)    &1.833(4)&1&$i$0.04(x)&&1.861(2)&2&0.035(2)&&1.8588(13)&2&0.047(16)&0.064(9)&&0.85(5)\\
Cu(1)-O(4)$_b$&1.941(4)&1&$i$0.01(x)&&-&0&-&&-&0&-&-&&-\\
Cu(2)-O(4)    &2.220(5)&1&0.04(x)&&2.264(3)&2&0.059(2)&&2.2684(15)&2&0.037(14)&0.064(9)&&0.6(1)\\
Cu(2)-O(4)$_b$&2.337(5)&1&$i$0.01(x)&&-&0&-&&-&0&-&-&&-\\
Cu(2)-Y&3.182(7)&8&0.029(6)&&3.191(8)&8&0.039(4)&&3.2041(5)&8&0.037(14)&0.034(16)&&0.44(10)\\
Cu(2)-Cu(2)&3.349(8)&8&0.10(2)&&3.358(8)&8&0.19(6)&&3.386(1)&8&0.037(14)&0.037(14)&&-3(2)\\
Cu(2)-Ba&3.382(15)&8&0.024(8)&&3.38(1)&8&0.037(8)&&3.367(1)&8&0.037(14)&0.046(16)&&0.60(8)\\
Cu(1)-Ba&3.475(8)&8&0.027(9)&&3.472(4)&8&0.037(8)&&3.462(1)&8&0.047(16)&0.046(16)&&0.55(11)\\
Cu(2)-O(2,3)&3.65(2)&2&0.12(1)&&3.656(3)&2&0.12(4)&&3.6583(5)&2&0.037(14)&0.056(17)&&-2(3)\\
Cu(1)-Cu(2)&4.102(5)&2&0.058(1)&&4.106(4)&2&0.064(1)&&4.1272(1)&2&0.047(16)&0.037(14)&&-0.03(10)\\
\end{tabular}
\label{thetable}
\label{table_fit}
\end{table}


\newpage
\onecolumn

\widetext

\newpage

\newpage

\newpage 

\newpage

\newpage

\newpage 

\end{document}